# Mid-infrared spectra of differentiated meteorites (achondrites): Comparison with astronomical observations of dust in protoplanetary and debris disks

Andreas Morlok[a,b,c]    a.morlok@open.ac.uk    *(Corresponding Author)*

Telephone: +44 1908 653219

Fax: +44 1908 858022

Chiyoe Koike[d]    koike-c@mua.biglobe.ne.jp

Kazushige Tomeoka[a]    tomeoka@kobe-u.ac.jp

Andrew Mason[c]    a.mason@open.ac.uk

Carey Lisse[e]    carey.lisse@jhuapl.edu

Mahesh Anand[c]    m.anand@open.ac.uk

Monica Grady[c]    m.m.grady@open.ac.uk

[a]Department of Earth and Planetary Sciences, Faculty of Science, Kobe University, Nada, Kobe 657-8501, Japan *(Affiliation Address)*

[b]Dept. of Mineralogy, The Natural History Museum, Cromwell Road, London SW7 5BD, UK

[c]Department of Physical Sciences, The Open University, Walton Hall, Milton Keynes MK7 6AA UK *(Present Address Corresponding Author)*

[d]Department of Earth and Space Science, Graduate School of Science, Osaka University, Toyonaka, Osaka 560-0043, Japan

[e]Space Department, Solar System Exploration Branch, Johns Hopkins University Applied Physics Laboratory, 11100 Johns Hopkins Road, Laurel, MD 20723, USA

**Abstract**

Mid-infrared (5 μm to 25 μm) transmission/absorption spectra of differentiated meteorites (achondrites) were measured to permit comparison with astronomical observations of dust in different stages of evolution of young stellar objects. In contrast to primitive chondrites, achondrites underwent heavy metamorphism and/or extensive melting and represent more advanced stages of planetesimal evolution. Spectra were obtained from primitive achondrites (acapulcoite, winonaite, ureilite, and brachinite) and differentiated achondrites (eucrite, diogenite, aubrite, and mesosiderite silicates). The ureilite and brachinite show spectra dominated by olivine features, and the diogenite and aubrite by pyroxene features. The acapulcoite, winonaite, eucrite, and mesosiderite silicates exhibit more complex spectra, reflecting their multi-phase bulk mineralogy.

Mixtures of spectra of the primitive achondrites and differentiated achondrites in various proportions show good similarities to the spectra of the few Myr old protoplanetary disks HD104237A and V410 Anon 13. A spectrum of the differentiated mesosiderite silicates is similar to the spectra of the mature debris disks HD172555 and HD165014. A mixture of spectra of the primitive ureilite and brachinite is similar to the spectrum of the debris disk HD113766. The results raise the possibility that materials produced in the early stage of planetesimal differentiation occur in the protoplanetary and debris disks.

# 1. Introduction

Dust in young stellar objects changes its composition and structure during the different stages of evolution of the objects (e.g., Apai and Lauretta, 2010). Those changes can be traced by studying characteristic emission features in the mid-infrared spectra of young main sequence (MS) stars and young proto-stellar objects (YSOs) (e.g., Lisse *et al.,* 2008). Comparison of those observations to laboratory measurements of dust analogs also provides information about the type of accreting planetesimals and physical and chemical processes occurring in the planetary systems. Because the planetesimals in those systems probably consist of materials similar to those in our Solar System, infrared measurements of meteorites provide an important source of information.

Most are complex assemblages of many kinds of minerals (Tab.1), and they cannot be reproduced simply by the combination of standard minerals. While we know major constituent minerals in the meteorites, exact modal abundances of all of the constituent minerals are not known, and many of the minerals are solid solutions, which means that they have variable compositions. Furthermore, individual meteorites have undergone various degrees of modification processes including alteration, metamorphism, and melting after the formation of their parent bodies. All of these effects should be reflected in many ways in measured mid-infrared spectra. However, we poorly know how these effects are related to infrared properties. Therefore, it is important to make direct mid-infrared measurements of the actual meteorites, material that underwent all these processes. This helps to estimate the impact of these processes on the mid-infrared spectra.

Previous laboratory mid-infrared measurements of meteorites as dust analogs mainly focused on primitive chondritic meteorites (e.g., Morlok *et al.,* 2006, 2008, 2010; Osawa *et al.,* 2001; Posch *et al.,* 2007; Zaikowski and Knacke, 1975). This study focuses on differentiated stony meteorites—achondrites. While the chondrites are pristine or only slightly processed from the primary nebular materials and thus represent a very early stage of planetesimal evolution, most of the achondrites underwent large-scale melting in the parent bodies and thus represent relatively more advanced stages of the evolution. Therefore, the detection of achondritic materials in circumstellar dust, if any, would help to identify the relatively evolved stages of disk formation.

The achondrites can be divided into two major groups—primitive achondrites and differentiated achondrites. The distinction between them is based on the degree of processing of the primary chondritic materials. Primitive achondrites are rocks that underwent heavy metamorphism with partial melting. Their compositions are moderately fractionated from the range of chondritic materials, which means that they are still somewhat similar to the solar composition. Differentiated achondrites are igneous rocks that underwent complete melting and recrystallization. Their compositions are highly fractionated, which means that they are very different from the solar composition due to differentiation during melting (e.g., McCoy *et al.,* 2006; Mittlefehldt, 2005).

Dusty exosystems of interest for this work are protoplanetary and debris disks. Protoplanetary disks represent the first stage of Solar System evolution, where the disks consist primarily of pristine dust and gas and materials partially processed by thermal events such as shock waves, X-ray flares, lightning, and the X-wind (Apai and Lauretta., 2010). Planetary formation (e.g., Greaves and Rice, 2010) and differentiation (e.g., Burkhardt *et al.*, 2008) are thought to begin very early in this protoplanetary disk stage. Through a transitional disk stage, the debris disk stage follows, in which pristine materials have been lost, and dust is almost entirely produced by collisions of planetesimals (e.g., Chen *et al.,* 2006; Lisse *et al.,* 2008, 2009, 2011; Morlok *et al.,* 2010). The age range for these disks overlaps with that for the formation of achondrites in our Solar System (Table 1). The protoplanetary disk phase lasts for less than ~10 million years, followed by the longer debris disk stage (Apai and Lauretta*,* 2010; Wyatt, 2007).

Here we present the results of mid-infrared laboratory room-temperature transmission/absorption spectroscopic measurements of primitive achondrites and differentiated achondrites. Previously, Sandford *et al*. (1984, 1993, 2010) performed comprehensive infrared transmission measurements of achondrites, but they mainly focused on differentiated achondrites and comparison with laboratory spectra of interplanetary dust particles and remote sensing spectra of asteroids. We obtained infrared spectra of the meteorites in the form of mass absorption coefficients. This form of spectra is valid for comparison with the emission of micron-size circumstellar dust, since the absorbance efficiency of a particle equals its emissivity in thermodynamic equilibrium (Posch, 2005). Our goal in this paper is: (1) to examine how the mineralogy of achondrites are related to infrared properties of those meteorites, and (2) to compare infrared spectra of the achondrites with infrared spectra of dust obtained from selected observations of young stellar objects, with an aim to form a basis for future detailed investigations.

## 2. Materials and Methods
### 2.1 Sample selection and preparation

The Smithsonian Institution (Washington), Institut für Planetologie (Münster), and the Natural History Museum (London) provided achondrite samples used for the present study (Table 1). Most of the samples from the Smithsonian Institution are aliquots of the standard materials used for bulk chemical analyses by Jarosewich (1966, 1990). To ensure that varying sizes of components, mineral grains, and fragments in meteorites do not introduce a sampling bias, an amount of at least 50 mg of a bulk meteorite sample was used to prepare a transmission pellet for each infrared measurement. For preparing pellets, bulk samples were ground to a very fine powder in an agate mortar or, in the case of metal-rich samples, a tungsten carbide mortar. 2 mg of meteorite powder was mixed with 1000mg KBr, with constant sample/KBr ratios in all cases. The mixture was ground for 75 minutes in the mortar, following the proven standard procedure (e.g. Chihara et al., 2002) to achieve grain sizes of <1 μm. We dried the powder in an oven for at least 24 hours at 110°C in air, to avoid additional water absorption. Finally, KBr-pellets were pressed on the powder for 45 minutes at 8t per cm$^2$.

**2.2 Infrared measurements**

We used a Nicolet Nexus 670 FT-IR spectrometer at the Department of Earth and Space Science, Osaka University, for the infrared measurements. Resolution of each measurement in a range from 2.5 μm to 25 μm was 4 cm$^{-1}$. 256 scans were summed up for each analysis. Blank KBr-pellets with the same thickness and mass as the sample pellets were used as references. For the absorbance data, the mass absorption coefficient (MAC) was calculated using κ=S/M ln (100/T), where S is the surface area of the pellet, M is the mass of sample in the pellet, T is the % transmission at a given wavelength, and κ is in (cm$^2$/g). Atmospheric $CO_2$ provided a constant reference feature at ~14.95 μm that we were able to use for spectral and absorbance calibration. For processing of the results, we used the ACD/SpecManager software. A lorenzian function was applied to obtain peak parameters, band positions, relative intensities, and widths (Full Widths at Half Height; FWHH).

For the comparison with the astronomical data, we used simple linear combinations of the laboratory spectra. Since the achondrites studied are complex assemblages of many kinds of minerals, detailed fittings are beyond the scope of our study. The aim of our comparison is to find general similarities between spectra of achondrites and dust in circumstellar environments. We calculated the mixtures of spectra 'by hand' based on first-order similarities of band positions. The percentages shown for each achondrite mixture are the weight proportions of individual achondrites, but given the uncertainties involved (see 2.3 Analytical Problems), the actual precision of the reported mixtures is on the order of magnitude. Comparisons of band positions are based on the values obtained by the peak fitting. If differences of band positions are within 0.2 μm, they are regarded as being similar (e.g., Morlok et al., 2008). Calculated mixtures (in Mass Absorbance Coefficient κ) and astronomical spectra in emissivity (continuum subtracted Flux Fv/Bv(T)) are normalized to unity.

The astronomical spectra for V410 Anon 13 (Sargent et al., 2006), HD172555 (Chen et al., 2006, Lisse et al., 2009), HD165014 (Fujiwara et al., 2010), and HD113766 (Lisse et al.,

2008) were obtained with the 5-35 um Infrared Spectrograph (IRS) on board the Spitzer Space Telescope. Sources for spectra of HD104237 are the 2.4 – 45 um Short Wavelength Spectrometer (SWS) on the Infrared Space Observatory (ISO) (Sloan et al., 2003) and the 8-13 um TIMMI2 spectrograph at the European Southern Observatory (ESO) (van Boekel et al., 2005).

**2.3 Analytical problems**

The spectra from the ureilite and eucrite are averages of representative spectra from three different meteorites, whereas the spectra from the other meteorites are representatives for individual meteorites (Table 1). The overall intensities of the features of measured spectra tend to vary considerably within the groups, while the band positions are constant. A major reason for these variations is probably the heterogeneous distribution of Fe-Ni metal, especially for the acapulcoite, winonaite, and mesosiderite samples. Sandford (2010) also observed similar effects of Fe-Ni metal in a series of infrared measurements of ureilites. Because Fe metal has a high density and strong absorbance without characteristic bands in the mid-infrared range, it easily influences the measurements by either producing a strong background or affecting the intensities of silicate bands. In addition, the material strength of metal makes it difficult to grind to a small size. Although relatively large metal grains were removed either by sieving the samples (Jarosewich, 1966) or by using a magnet (this study), usually some metal remained. Due to high contents of remaining metal, most spectra from mesosiderite samples have only very low intensities. Thus, we selected only the mesosiderite spectrum with the highest intensity, whose band positions are similar to the other spectra, which are also shown in Fig. 1c. We also selected the eucrite spectra with the highest intensities for the same reason. From the acapulcoite and winonaite samples, only comparatively weak spectra were obtained, which probably resulted from the presence of remaining metal.

Fe-Ni sulfides may also affect the spectra in a manner similar to that of Fe metal. This problem has been recognized in the earlier studies of ureilites and other achondrites (Nittler *et *al.,* 2004; Sandford *et al.,* 1984, 2010). Acapulcoites contain up to 11 vol%, winonaites up to 28 vol%, and mesosiderites up to 12 vol% Fe-Ni sulfides (Kimura *et al.,* 1992).

Ureilites contain significant amounts of carbon (2–6 wt%, Mittlefehldt, 1998). This phase cannot be separated from the silicates, and its presence explains the strong underlying featureless continuums of the spectra. Sandford et al. (2010) observed a similar effect. Thus, we discarded spectra of ureilites with significantly low intensities (below 3000 κ).

**3. Mineralogy of Achondrites**

In this section, to help readers understand the following descriptions, we provide a brief summary of the mineralogy of achondrites used for our infrared measurements as well as information regarding their parent asteroids. Because of the rather homogeneous nature of the achondrites and the large amounts of the samples used, we mostly depend on the published literature (e.g., Hutchison, 2004; Mittlefehldt *et al.,* 1998; Mittlefehldt, 2005) for their mineralogy and petrology. The mineralogy and petrology of the samples are also summarized in Table 1.

Previous studies suggested that silicate melting on the achondrite parent bodies probably started at ~1323 K and that substantial melting occurred in the temperature range between 1500 and 1800 K (e.g., McCoy *et al.,* 2006).

Among the primitive achondrites, the acapulcoites and winonaites have essentially primitive chondritic mineralogy with ultramafic compositions. Orthopyroxene ($En_{99-86}$), forsteritic olivine ($Fo_{99-86}$), and albitic plagioclase ($An_{31-8}$) dominate these meteorites. Both acapulcoites and winonaites also contain minor proportions of Ca-rich pyroxene, Fe-Ni metal and Fe sulfide (troilite). Reflectance spectroscopic studies indicated that acapulcoites and winonaites have similarities to S (IV) and S (VII) type asteroids (Gaffey *et al.*, 1993; Hiroi *et al.*, 1993; McCoy *et al.*, 1997).

The ureilites and brachinites have been classified as primitive achondrites (e.g., Weisberg *et al.,* 2006). However, they also have some features suggesting a highly fractionated origin (e.g., Mittlefehldt, 2005). Thus, one has to keep in mind that these meteorites may exhibit characteristics of both primitive and differentiated achondrites. The ureilites consist dominantly of olivine (Fo$_{95-75}$), orthopyroxene (En$_{90-80}$), and Ca-rich pyroxene (Wo$_{15}$) in various proportions, with minor elemental carbon. The reflectance spectral features of ureilites are most similar to C-class asteroids (Clouts *et al.*, 2010). Ureilites are the only achondrite group so far directly and physically linked to an extraterrestrial body, asteroid 2008 TC3 (Jenniskens *et al.*, 2010). The brachinites consist mainly of olivine (Fo$_{70-65}$) and Ca-rich pyroxene, with minor variable amounts of albitic plagioclase (An$_{32-22}$). The A-class asteroid 289 Nenetta has been identified to be similar to brachinites (Sunshine *et al.*, 2007).

The differentiated achondrites considered here have lower olivine contents compared with the primitive achondrites. The eucrites mainly consist of plagioclase (An$_{98-60}$) and Ca-rich pyroxene. The diogenites are composed dominantly of orthopyroxene (enstatite), with significant proportions of plagioclase (An$_{91-60}$) and olivine (Fo$_{73-65}$). Spectroscopic data favor identification of 4 Vesta as the parent body of the eucrites and diogenites (Consolmagno and Drake, 1977; McCord *et al.*, 1970).

The aubrites consist mainly of orthopyroxene and Ca-rich pyroxene (Wo$_{50}$), with some forsteritic olivine. The E type or rare Xe-type asteroids have been suggested to be among the parent objects for aubrites (Binzel *et al.*, 2004; Burbine *et al.*, 2001, De Luise *et al.*, 2007; Gaffey *et al.*, 1992; Nedelcu *et al.*, 2007).

Mesosiderites are essentially stony irons. The sample used for the present study is a non-metallic fraction of a mesosiderite, which is a mixture composed mainly of Mg-rich olivine (Fo$_{92-58}$), Ca-rich pyroxene, orthopyroxene, and plagioclase; this sample will be referred to as mesosiderite silicates. Mesosiderites are considered to have been derived from a large asteroid, 200–400 km in diameter, with a potential link to the HED parent body (Greenwood *et al.*, 2006; Haack *et al.*, 1996; Mittlefehldt, 2005; Scott *et al.*, 2001).

The various types of achondrites can show significant variations between different members of the same group (e.g. Mittlefehldt *et al.,* 1998). However, it is impossible to cover all variations in the scope of this study. Therefore, in this study we focus on typical samples for the achondrite groups (Table 1).

[Here should be Figures 1a-c]

## 4. Results

The mid-infrared spectra measured from various groups of achondrites are shown in Figs. 1a–c. Wavelengths of major bands of the measured spectra are shown in Table 2.

### 4.1 Primitive achondrites

The acapulcoite and winonaite show very similar spectra (Fig. 1a, Table 2). Both meteorites show strong bands with minor differences in relative intensity at 9.4–9.5, 9.9, 10.2, 10.8, and 11.3 μm. They also show further strong bands at 19.8 μm and 23.9 μm and smaller bands at 13.8 μm and 14.5 μm. All of these features can be explained by orthopyroxene and Ca-rich pyroxene. Most of the characteristic features for olivine and plagioclase are probably affected by the pyroxene bands; thus, it is difficult to specify those features (Chihara *et al.,* 2002; Koike *et al.,* 2000; Kimura *et al.,* 1992; Yugami *et al.,* 1998).

The spectrum of ureilite is characterized by strong bands at 9.3, 10.1, 11.2, 16.4, 19.6, and 23.8 μm (Fig. 1a, Table 2), all of which are basically olivine features except the band at 9.3 μm which is typical of orthopyroxene (Koike *et al.,* 2000, 2003; Morlok *et al.,* 2006). The spectra of individual ureilites used to calculate the averaged spectrum show high similarities in positions, intensities, and shapes of bands. This confirms the general homogeneity of this group of meteorites. The results are comparable to those of the other study of ureilites (Sandford *et al*., 2010), although there are rare cases of pyroxene-rich samples (Sandford, 1993). The averaged spectrum of multiple samples from the Almahata Sitta meteorite (a unique polymict ureilite recovered from a recent fall) (Sandford *et al.,* 2010) is very similar to the ureilite spectrum of this study.

The spectrum of brachinite has an overall resemblance to that of the ureilite with strong features at 10.1, 11.3, 16.9, ~20, and 23.9 μm (Fig. 1a, Table 2), all of which are also typical olivine bands (Koike *et al.*, 2003; Morlok *et al.*, 2006). The small features between 12.5 μm and 13.0 μm can be ascribed to the presence of crystalline silica, which is not common in brachnites; this may be a contamination from other material.

**4.2 Differentiated achondrites**

The eucrite shows a spectrum characterized mainly by a strong double feature at 10.5–10.7 μm, two further features at 9.3 μm and 9.7 μm, and a broader feature at ~20 μm (Fig. 1b). Further small features are present at 11.2, 13.7, 16.0, 17.6, 18.5, and 24 μm. The shape of the main feature between 8 μm and 12 μm is characteristic of anorthite (Salisbury *et al.*, 1991). The spectra of the individual eucrites used to calculate the averaged spectrum are very similar to each other, indicating a high degree of intra-group bulk homogeneity. The spectrum is similar to those of the Juvinas and Nobleborough eucrites reported by Sandford (1984), indicating that these features are indeed typical of this large group.

The spectrum of the diogenite shows strong bands at 9.4, 10.6, 11.5, and 19.9 μm (Fig. 1b), which are typical features of orthopyroxene (Koike *et al.*, 2000). Additional weaker bands are found at 13.4, 13.8, 14.6, 15.5, 18.7, 21.9, and 22.4 μm.

The spectrum of the aubrite has an overall similarity to that of the diogenite with strong bands at 9.3, 9.9, 10.7, 11.7, 19.6, and 20.7 μm (Fig. 1b, Table 2). Weaker features occur at 13.8 μm, 18.5 μm, and 21.9 μm. All of them are typical of orthopyroxene spectra (Bowey *et al.*, 2007; Koike *et al.*, 2000).

The mesosiderite silicates show a spectrum with a dominating feature at 9.13 μm (Fig. 1c, Table 2), which can be attributed to an overlap of the features for pyroxene (Chihara *et al.*, 2002) and a $SiO_2$ phase (Koike *et al.,* 1994) that constitutes up to ~15 vol% of the mesosiderite silicates (Delany *et al.,* 1981; Kimura *et al.,* 1991). Other strong bands are present at 10.6 μm, 11.3 μm, and 20.3 μm, which mainly resulted from orthopyroxene and Ca-rich pyroxene (Koike *et al.,* 2000). Mesosiderites are very heterogeneous rocks consisting of fragments of different sizes, shapes, and compositions (Mittlefehldt *et al.,* 1998). However, the Patwar meteorite sample used in this study is an aliquot from a large amount of processed samples (>9 g, Jarosewich 1966, 1990); thus, it probably gives a reasonable average of various fragments. The weaker spectra of the other three mesosiderites (Woodbine, Barea, Emery) presented in Figure 1c also show high general similarities to the spectrum of Patwar.

**5. Discussion: Comparison to Astronomical Spectra**

In this section, we compare the spectra from the achondrites with the spectra from astronomical measurements of selected protoplanetary disks and debris disks (Figs. 2a–b; Table 3). The astronomical spectra are measured with both ground-based and space-based instruments. Our comparison is focused on the spectral range from 5 μm to 25 μm. Some of the astronomical spectra measured by ground-based instruments are limited to the narrow atmospheric window from 8 μm to 13 μm. All the astronomical spectra have their continuum subtracted and are presented in Emissivity, in order to allow a better comparison with the Mass Absorption Coefficients. In the following comparison, we mainly focus on the major spectral features (position, intensity, shape, and width of bands) indicating the composition and mineralogy of the observed samples and dust. Laboratory measurements and calculations indicate that the infrared features also depend on various factors including size, shape, structure, and temperature of observed materials (e.g., Bowey et al., 2001; Min et al., 2005). In the following discussion, we do not go into details of these factors and mainly focus on a first-order approximation to the astronomical spectra to deduce potential objects that constitute the dust.

**[Here should be Figures 2a-b]**

**5.1 Protoplanetary Disks**

V410 Anon 13 is a T Tauri star (spectral type M6) star with an assumed age of 1-3 Myr (Fig. 2a) (Furlan *et al.,* 2006; Muzerolle *et al.,* 2000; Sargent *et al.,* 2006; White and Ghez, 2001). Its spectrum is similar in band position to a mixture of the spectra of winonaite (38%), brachinite (38%), and mesosiderite silicates (24%). The dominant feature at 9.4 μm and weaker features at 10.1 μm, 10.3 μm, 10.8 μm, and 11.3 μm in the astronomical spectrum find corresponding features within 0.2 μm of each peak position in the laboratory spectrum. Since the astronomical spectrum is very complicated with many fine features, precise identification of several weaker bands up to 11.3 μm is difficult. However, the laboratory spectrum is similar in relative intensity of the silicate feature in the range between 9 μm and 12 μm. In addition, a broad ~20-μm feature similar to the range between 16 μm and 23 μm in the astronomical spectrum has been generated.

HD104237A is a Herbig star (spectral type A8Ve) with an assumed age of 5 Myr (Sloan *et al*., 2003; Van Boekel *et al*., 2005). Van Boekel et al. (2005) suggested that the dust around this star is largely composed of amorphous material with a pyroxene composition. We found that a mixture of the spectra of acapulcoite (40%), brachinite (40%), and mesosiderite silicates (20%) is similar to the astronomical spectrum of this star (Fig. 2a). The three strong bands at 9.3 μm, ~10.1 μm and ~11 μm in the astronomical spectrum find equivalents within 0.2 μm in the silicate features at 9.2 μm, 10.0 μm and 11.3 μm in the laboratory spectrum. In addition, a broad feature at ~20 μm occurs in both the laboratory and dust spectra.

These general similarities between the spectra of the achondrite mixtures and the young stellar objects suggest that achondritic materials occur in addition to pristine and/or partially processed primitive materials in some of the protoplanetary disks, pending more detailed studies with a wider range of materials. Chronological investigations indicate that the ages of most achondrites are close to that of our Solar System (4567 ± 0.1 Ma based on the age of a CAI; Amelin *et al.*, 2002). At least the maximum ages of ureilites, brachinites, basaltic eucrites, and mesosiderite silicates overlap or fall within 1 Ma of this point (Tables 1 and 3; Elkins-Tanton *et al.,* 2011), and those of the other achondrites fall within 10 Ma, with the exception of winonaites (Table 1). These ages except for that of winonaites overlap with the age range (1 to 5 Ma) of the protoplanetary systems used in this study (Table 3). Chronological constraints obtained so far suggest that differentiation of planetesimals starts very early (possibly within the first few million years) in the evolution of the disks (e.g., Burkhardt *et al.*, 2008; Wadhwa *et al.,* 2006). A prerequisite for the presence of achondritic dust in such an early stage is the occurrence of frequent collisions in the gas-rich protoplanetary disks. Such collisions can be expected even in optically thick circumstellar disks (Lisse *et al.,* 2009).

However, the following caveat is necessary in the interpretation based on the infrared measurements. The complex structure of an optically thick protoplanetary disk commonly hinders obtaining clear mid-infrared information from the actual planetesimal-forming regions. For example, infrared spectra may be significantly affected by processes such as mixing of materials caused by disk turbulence, radiation transfer, and introduction of materials from other sources (e.g., Pontoppidan and Brearley, 2010; Van Boekel *et al*., 2005). More detailed astronomical observations would be required for solution of these problems. Furthermore, these comparisons are yet based on a limited selection of samples. To remove ambiguities, more systematic comparisons of a broader range of samples are necessary.

**5.2 Debris Disks**

HD 165014 is a main-sequence F2V star with a circumstellar disk of dust between 0.7 AU and 4.4 AU; its precise age is unknown (Table 3). The dust spectrum is very similar to that of HD 172555 (Fig. 2b), with a strong band at 9.3 μm and a characteristic band at 10.4 μm with a shoulder at 11.0 μm. There is also a very broad hump between 14 μm and 22 μm with a maximum at 18 μm (Fujiwara *et al.,* 2010). The first three features resemble the three characteristic bands of the spectrum from the mesosiderite silicates at 9.1 μm, 10.6 μm, and 11.3 μm, with only small shifts (0.2 μm and less). The main difference between the astronomical and the laboratory spectra is the presence of a broad, strong feature at ~19 μm in the former (and at ~20 μm in the mesosiderite silicates). Fujiwara *et al.* (2010) modelled HD 165014 as a mixture of enstatite with small amounts of olivine and amorphous silica; crustal stripping from a Mercury-type planet was suggested as a possible source for the dust.

HD172555 is a ~12 Myr old A5V star (Lisse *et al.*, 2009; Chen *et al.*, 2006; Zuckerman and Song, 2001) with a warm debris disk. A strong, broad feature at 9.4 μm (Fig. 2b) characterizes the spectra from dust around both stars. Among the achondrites, the mesosiderite silicates also exhibit a similar shape spectrum. However, the strong pyroxene bands at 10.6 μm and 11.3 μm do not occur in the spectrum of HD172555, suggesting a lack of pyroxene in this star. Lisse *et al.* (2009) interpreted that the disk of HD172555 mainly consists of a silica dust-SiO vapour mixture produced by hypervelocity collisions of large differentiated planetesimals. Our results support this interpretation, since mesosiderites are considered to be debris or polymict breccia resulting from the collisions of differentiated planetesimals (Hutchison, 2004; Mittlefehldt, 2005), which possibly have grown up to 200–400 km-diameter bodies (Scott *et al.*, 2001).

HD113766A (spectral type binary F3/F5V system, assumed age 10–16 Myr) (Chen *et al.*, 2006; Lisse *et al.*, 2008; Mamajek *et al.,* 2002) also has a warm circumstellar debris disk. A mixture of the spectra of brachinite (66%) and ureilite (34%) produces a spectrum similar to that from HD113766A (Fig. 2b). In the laboratory spectrum, the strong bands corresponding to the features at 10.0 μm and 11.1 μm and weaker features at 9.4 μm and 11.9 μm in the astronomical spectrum appear slightly shifted (less than 0.2 μm) to higher wavelengths. These differences may be ascribed to variable Fe/Mg contents in olivine (e.g. Koike *et al,* 2003). This interpretation is consistent with the suggestion of Lisse *et al.* (2008) that ureilites (which contain major proportions of olivine) are the dominant component of the dust in HD113766A.

Based on their visible spectroscopy/near-infrared spectra, the olivine-rich brachinites have been matched to A-type asteroids (Sunshine *et al.*, 2007) and ureilites to S-type asteroids (Gaffey *et al.*, 1993) in our Solar System. If this is the case, then the match between the infrared spectra of the brachinite-ureilite mixture and HD 113766A implies that the debris disk has arisen through collisions between planetesimals similar to the A-type and S-type asteroids. Such collisions are also possible within our own asteroid belt, since both A-type and S-type asteroids are Main Belt objects, with mean orbits between 1.8 and 3.2 AU (JPL Small Body Database Search Engine, accessed 31st July 2011; http://ssd.jpl.nasa.gov/sbdb_query.cgi)

## 6. Conclusions

Mid-infrared spectra of the achondrites show a wide range of spectral features reflecting the various mineralogies of the meteorites. Pending further detailed studies with a wider range of materials, our results suggest that the achondritic materials occur in addition to pristine and/or partially processed primitive materials in the protoplanetary and debris disks.

The mixtures of spectra from primitive and differentiated achondrites in various proportions show good similarities to the astronomical spectra from dust in the protoplanetary disks. This would support the model that differentiation of planetesimals starts very early in the evolution of the disks (e.g., Burkhardt *et al*., 2008; Wadhwa *et al.,* 2006) and that the formation of dust by the collisions of differentiated planetesimals occurs in this early stage.

The spectra from mesosiderite silicates and a brachinite/ureilite mixture show particularly good fits with the astronomical spectra from dust in the debris disks. Mesosiderites are regarded as chaotic mixtures of debris produced by collisions of highly differentiated bodies; thus, the results further support the proposal that formation of dust by collisions of highly differentiated planetesimals occurs in this stage of disk evolution.

**Acknowledgements**

Many thanks to Linda Welzenbach (Smithsonian Institution, Washington), Caroline Smith (The Natural History Museum, London), and Addi Bischoff (Institut für Planetologie, Münster) for the samples. We thank Karly Pitman and an anonymous reviewer for constructive and helpful reviews. This work was supported by "The 21st Century COE program of Origin and Evolution of Planetary Systems" of the Ministry of Education, Culture, Sports, Science and Technology at Kobe University (to A.M. and K.T.), and the Grant-in-Aid for Scientific Research (No. 20340150 to K.T.); PPARC (A. M.; M. G.). Also thanks to J.P. Mason (OU) for helping with the spectra. Many thanks for providing the astronomical spectra of this paper to Roy van Boekel (HD104237A), C. Chen (HD172555), H. Fujiwara (HD165014), C. Lisse (HD113766A), B.A.Sargent (V410 Anon 13). This work is partially based on observations made with the Spitzer Space Telescope, which is operated by the Jet Propulsion Laboratory, California Institute of Technology, under NASA Contract 1407. Part of this work is based on observations with ISO, an ESA project with instruments funded by ESA Member States (especially the PI countries: France, Germany, The Netherlands, and The United Kingdom) and with the participation of ISAS and NASA. C. Lisse gratefully acknowledges support from JPL contract 1274485, the APL Janney Fellowship program, and NSF Grant AST-0908815.

Table 1. Samples used in this study and their petrologies, mineralogies, and ages. BM=The Natural History Museum, London; Mue=Institut fuer Planetologie, Muenster; USNM=Smithsonian Institution, Washington. **BOLD:** Major components (>5 vol%). Data are from various literature sources. References: (for Petrology and Mineralogy) mainly Hutchison (2004), and additional information from Mittlefehldt *et al*. (1998); Mittlefehldt (2005); For ages: [1] Stewart B. *et al.* (1996), [2] Terribilini D. *et al.* (2000), [3] Torigoye-Kita *et al.* (1995), [4] Kita N.T. *et al.* (2007), [5] Amelin Y. e*t al.* (2008), [6] Spivak-Birndorf L.J. & Wadhwa M. (2009), [7] Smoliar (1993) [8] Lugmair and Shukolyukov (1997), [9] Gilmour J. D. *et al.* (2006), [10] Ireland and Wlotzka (1992).

| Meteorite | Petrology | Mineralogy | Age (Ma) | Samples Analyzed |
|---|---|---|---|---|
| **Primitive** | | | | |
| Acapulcoite | Ultramafic, Chondritic Metamorphism, Partial Melting | **Orthopyroxene$_{(En97-86)}$, Ca-Pyroxene$_{(Wo44/En51)}$, Plagioclase$_{(An31-12)}$, Olivine$_{(Fo97-86)}$,** Kamacite, Taenite, Troilite, Phosphate, Chromite [Ol/Pyx ≤1] | 4557± 2 [2] | Dhofar 125 (Mue) |
| Winonaite | Ultramafic, Chondritic Metamorphism, Partial Melting | **Orthopyroxene$_{(En99-91)}$, Plagioclase$_{(An25-8)}$, Olivine$_{(Fo99-92)}$,** Kamacite, Taenite, Troilite, Daubreilite, Oldhamite, Alabandite, Phosphate, Schreibersite, Graphite, Chromite, Ca-Pyroxene [Ol/Pyx <1] | 4530± 20 [1] | Winona (USNM7073) |
| Ureilite | Ultramafic Cumulates Partial Melting Residue | **Olivine$_{(Fo95-75)}$, Orthopyroxene$_{(En90-80)}$, Ca-Pyroxene$_{(Wo15/En75)}$,** Kamacite, Troilite, Daubreilite, Halite, Phosphate, Schreibersite, Graphite, Chromite [Ol/Pyx >1] | 4563± 30 [3, 4] | Dingo Pup Donga (USNM7073) Novo Urei (USNM 2969) Goalpara (USNM1545) |
| Brachinite | Chondritic, Partial Melting Residue, Ultramafic Metamorphism/ Melt Crystallization | **Olivine$_{(Fo70-65)}$, Plagioclase$_{(An32-22)}$,** Ca-Pyroxene, Orthopyroxene, Taenite, Troilite, Phosphate, Chromite [Ol/Pyx >>1] | 4564± 2 [5, 6] | Nova 003 BM1993,M11 |
| **Differentiated** | | | | |
| Eucrite | (Ultra)Mafic Basaltic (B), or Cumulate Melt (C) | **Plagioclase$_{(An98-60)}$, Ca-Pyroxene** Silica, Kamacite, Troilite, Phosphate, Chromite, Orthopyroxene [Ol/Pyx <<1] | 4557−4566 (B) 4399-4484 (C) [7, 8] | Juvinas (USNM 2844) Moore County (USNM929) Nuevo Laredo (USNM1783) |
| Diogenite | (Ultra)Mafic, Melt Crystallization Cumulates | **Orthopyroxene$_{(En77-67)}$, Plagioclase$_{(An91-60)}$, Olivine$_{(Fo73-65)}$,** Ca-Pyroxene, Silica, Kamacite, Troilite, Phosphate, Chromite [Ol/Pyx <<1] | 4561 [8] | Bilanga (Mue) |
| Aubrites | Ultramafic Melt Crystallization/ Melt Residue | **Orthopyroxene$_{(En100)}$, Ca-Pyroxene$_{(Wo50/En50)}$, Olivine$_{(Fo100)}$,** Plagioclase, Kamacite, Taenite, Troilite, Daubreillite, Oldhamite, Phosphate, Chromite [Ol/Pyx <<1] | 4563 ± 1 [9] | Pena Blanca Springs (Mue) |
| Mesosiderite Silicates | Impact Melt/Breccia | **Ca-Pyroxene, Orthopyroxene, Plagioclase, Olivine, SiO$_2$,** Chromite, Phosphates, Ilmenite, Metal and Sulfide Clasts | 4653±15 [10] | Patwar (USNM7073) |

Table 2. Spectral parameters of the achondrite spectra (in μm). Peak = Band position in spectrum, Barycenter = Band position in fitted feature; FWHH = Peak size at Full Width, Half Height (in μm). Intensity: VS = very strong (90%-100% of maximum height), S = strong (60%-90%), M = medium (30%-60%), W = weak (10%-30%), VW = very weak (0%-10%).

| Peak | Barycenter | FWHH | Intensity | Peak | Barycenter | FWHH | Intensity |
|---|---|---|---|---|---|---|---|
| **Acapulcoite** | | | | **Eucrite** | | | |
| 8.92 | 8.74 | 1.40 | S | 8.19 | 8.29 | 2.77 | W |
| 9.39 | 9.38 | 0.99 | S | 9.34 | 9.26 | 0.70 | S |
| 9.93 | 9.90 | 0.34 | S | 9.71 | 9.71 | 0.57 | S |
| 10.21 | 10.21 | 0.57 | S | 10.50 | 10.41 | 0.51 | VS |
| 10.80 | 10.72 | 0.78 | VS | 10.65 | 10.71 | 0.54 | VS |
| 11.30 | 11.41 | 1.32 | VS | 11.15 | 11.27 | 0.98 | S |
| 13.83 | 13.84 | 0.32 | W | 13.19 | 13.12 | 0.61 | VW |
| 14.53 | 14.47 | 0.29 | VW | 13.72 | 13.73 | 0.48 | W |
| 19.79 | 19.79 | 1.91 | VS | 14.94 | 14.89 | 0.57 | W |
| 23.90 | 24.43 | 2.62 | VS | 16.00 | 15.96 | 0.85 | M |
| | | | | 17.58 | 17.36 | 1.77 | M |
| **Winonaite** | | | | 18.52 | 18.49 | 1.09 | M |
| 8.88 | 8.99 | 1.25 | M | 20.02 | 19.65 | 2.10 | M |
| 9.48 | 9.47 | 0.72 | S | 24.01 | 24.23 | 3.03 | W |
| 9.91 | 9.90 | 0.36 | S | | | | |
| 10.19 | 10.20 | 0.43 | S | **Diogenite** | | | |
| 10.78 | 10.66 | 0.86 | S | 8.96 | 8.90 | 0.29 | M |
| 11.25 | 11.40 | 1.53 | S | 9.41 | 9.45 | 0.46 | S |
| 13.83 | 13.79 | 0.51 | W | 10.58 | 10.57 | 0.31 | S |
| 14.52 | 14.51 | 1.38 | W | 11.45 | 11.47 | 1.07 | S |
| 15.62 | 15.51 | 1.33 | M | 13.79 | 13.78 | 0.35 | W |
| 18.45 | 18.36 | 1.41 | S | 14.57 | 14.56 | 0.42 | W |
| 19.79 | 19.72 | 2.01 | VS | 15.53 | 15.51 | 0.58 | W |
| 21.88 | 21.46 | 2.34 | S | 18.65 | 18.65 | 0.40 | S |
| 23.90 | 23.92 | 0.99 | VS | 19.87 | 19.62 | 1.57 | VS |
| 24.58 | 24.79 | 1.28 | S | 21.88 | 22.17 | 1.89 | M |
| | | | | | | | |
| **Ureilite** | | | | **Aubrite** | | | |
| 8.18 | 8.02 | 2.50 | W | 8.85 | 8.80 | 0.55 | M |
| 9.34 | 9.30 | 0.66 | M | 9.34 | 9.39 | 0.35 | S |
| 10.13 | 10.14 | 0.41 | S | 9.88 | 9.87 | 0.29 | S |
| 11.22 | 11.25 | 0.36 | VS | 10.71 | 10.73 | 0.53 | VS |
| 16.41 | 16.38 | 0.92 | W | 11.68 | 11.70 | 0.54 | S |
| 19.57 | 19.65 | 1.35 | S | 13.79 | 13.81 | 0.29 | W |
| 23.79 | 24.79 | 8.22 | S | 14.44 | 14.42 | 0.39 | W |
| | | | | 14.65 | 14.64 | 0.33 | VW |
| **Brachinite** | | | | 15.43 | 15.45 | 0.40 | W |
| 10.11 | 9.99 | 0.63 | VS | 17.64 | 17.56 | 0.57 | M |
| 10.58 | 10.61 | 0.29 | S | 18.52 | 18.39 | 0.82 | S |
| 11.27 | 11.29 | 0.61 | VS | 19.57 | 19.59 | 1.02 | VS |
| 11.95 | 11.96 | 0.44 | M | 20.74 | 20.75 | 0.60 | S |
| 12.50 | 12.50 | 0.47 | W | 21.88 | 21.88 | 0.83 | S |
| 16.89 | 16.73 | 1.47 | M | 23.05 | 23.20 | 1.08 | S |
| 19.35 | 19.05 | 2.45 | VS | 24.69 | 24.76 | 0.93 | M |
| 19.94 | 20.26 | 1.72 | VS | | | | |
| 21.79 | 21.63 | 1.81 | S | **Mesosiderite** | | | |
| 23.90 | 24.22 | 4.61 | S | 9.13 | 9.10 | 0.41 | VS |
| | | | | 10.56 | 10.58 | 0.67 | M |
| | | | | 11.30 | 11.29 | 0.35 | M |
| | | | | 12.62 | 12.51 | 1.07 | VW |
| | | | | 13.75 | 13.78 | 0.32 | VW |
| | | | | 14.69 | 14.75 | 0.72 | VW |
| | | | | 15.62 | 15.49 | 0.34 | W |
| | | | | 18.72 | 18.24 | 2.10 | M |
| | | | | 20.26 | 20.60 | 2.03 | W |

Table 3. Overview of the protoplanetary and debris disk systems used for comparison and the achondrites showing spectral matches.

| Star | Type | Age (Ma) | Ref | Spectral match (this study) |
|---|---|---|---|---|
| *Protoplanetary disks* | | | | |
| V 410 Anon 13 | M6 | 2.3 | 3 | Winonaite (38%) + brachinite (38%) + mesosiderite (24%) |
| HD 104237A | A8Ve | 5 | 1, 2 | Acapulcoite (40%) + brachinite (40%) + mesosiderite (20%) |
| *Debris disks* | | | | |
| HD 172555 | A5V | 12 | 6 | Mesosiderite |
| HD 165014 | F2V |  | 5 | Mesosiderite |
| HD 113766A | F3/F5V | 16 Ga | 4 | Brachinite (66%) + ureilite (34%) |

**References:** [1] Sloan *et al.,* 2003, [2] Van Boekel *et al.,* 2005, [3] White and Ghez, 2001, [4] Mamajek *et al.,* 2002, [5] Fujiwara *et al.,* 2010, [6] Zuckerman and Song, 2004

Figures Captions

Figures 1a–c: Infrared absorption spectra of achondrites. κ is Mass Absorption Coefficients (MAC) in cm$^2$/g. The averaged spectra of several single measurements are in continuous lines, the spectra of the individual component sin dotted lines. The spectra are shifted for clarity, number in brackets show the offset in κ (Mass Absorbance Coefficient). The similarity in band position and intensity between the single spectra of ureilites, eucrites and mesosiderites demonstrates the strong homogeneity of the sample materials.

Figure 2a: Comparison of the astronomical spectra (thick, continuous black lines) of older dusty circumstellar disks V410 (Furlan *et al*, 2006; Sargent *et al.,* 2006) HD104237A (sources: Sloan *et al.,* 2003; van Boekel *et al*., 2005) with mixtures of spectra of achondrites (light gray, continous line overlaying astronomical spectrum). The thin lines on the bottom are the single constituent of the calculated mixtures. These mixtures of winonaite/acapulcoite, brachinite, and mesosiderite show similarity in band position and shape with the astronomical data. The strongest features in the laboratory mixtures fall within 0.2 μm of the equivalent bands in the astronomical spectra, which indicates a significant similarity in mineralogical composition. κ is mass absorption coefficients (MAC) in cm$^2$/g; Fν/Bν(T) is emissivity, the astronomical flux divided by the blackbody at the dust temperature. The spectra of the laboratory mixtures and the astronomical spectra were normalized to unity, the same intensity based on the strongest feature in the 8-13 micron range to allow better comparison. Similar, the spectra of the single components are also normalized to the same intensity.

Figure 2b: Comparison of the astronomical spectra of dust-rich debris disks HD172555 (Chen *et al*., 2006; Lisse *et al*., 2009), HD165014 (Fujiwara *et al*., 2010) and HD113766A (Chen *et al*., 2006) (thick, continuous black lines). Calculated mixtures used for comparison are light gray, continuous lines overlaying the astronomical spectrum. Mesosiderite silicates show a good similarity (difference less than 0.2 μm) regarding the bands in the strong main feature around ~9.1 μm in comparison to the spectrum of HD165014. HD113766 is similar to a mixture of ureilites and brachinite. κ is Mass Absorption Coefficients (MAC) in cm$^2$/g; Fν/Bν(T) is emissivity, the astronomical flux divided by the blackbody at the dust temperature. The spectra of the laboratory mixtures and the astronomical spectra were normalized to unity, the same intensity based on the strongest feature in the 8-13 micron range to allow better comparison. Similar, the spectra of the single components are also normalized to the same intensity**.**

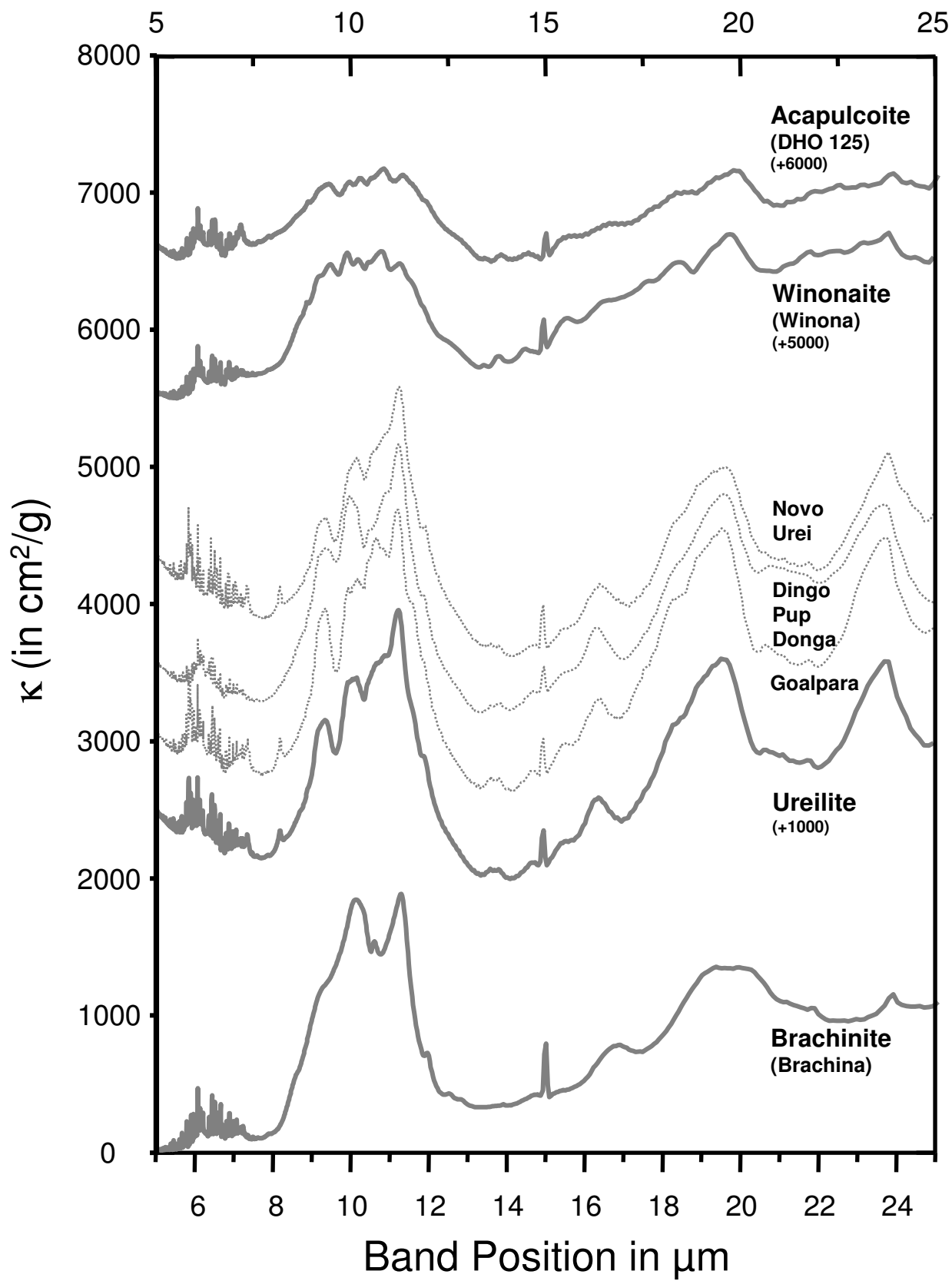

Figure 1a

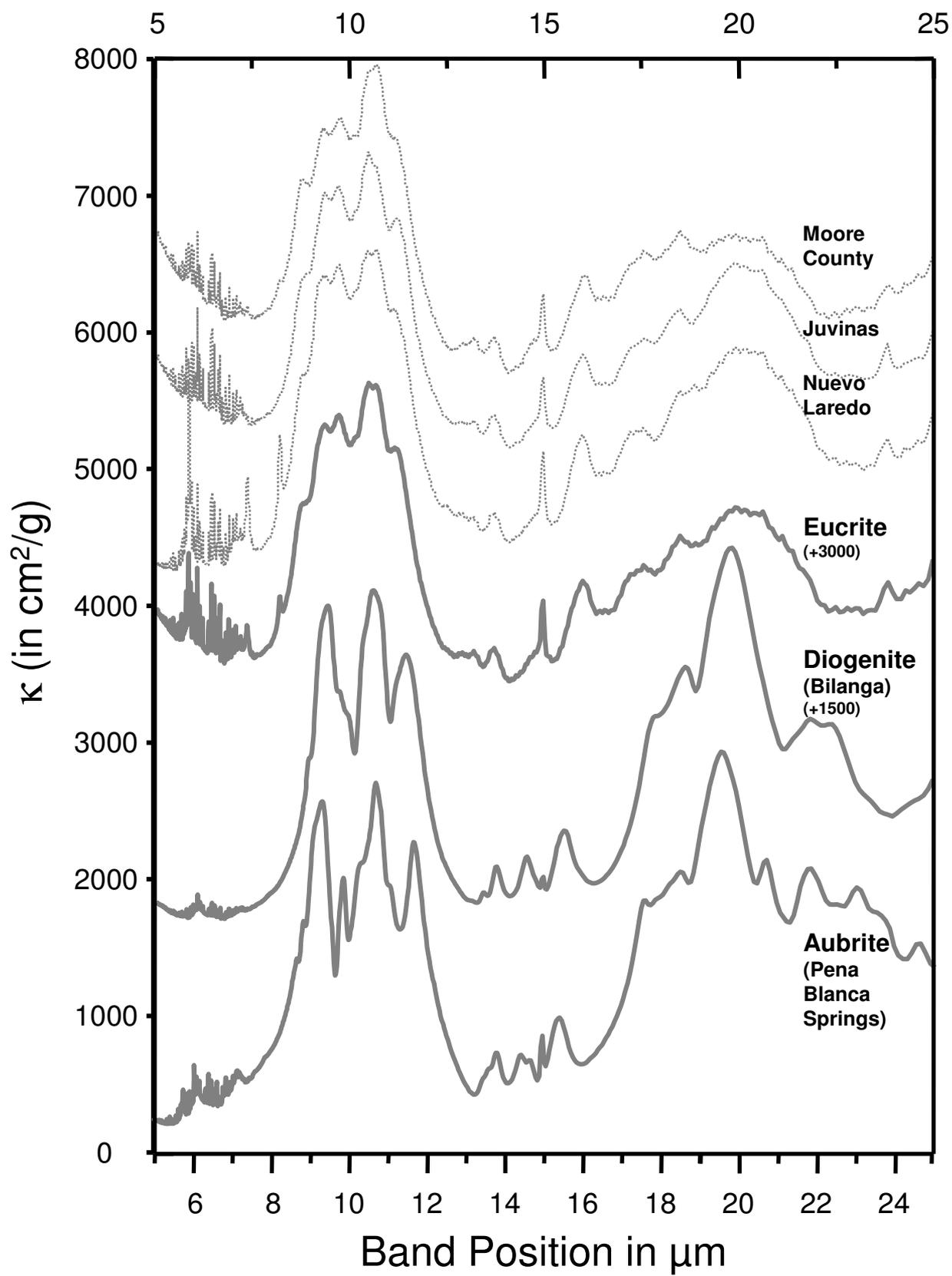

Figure 1b

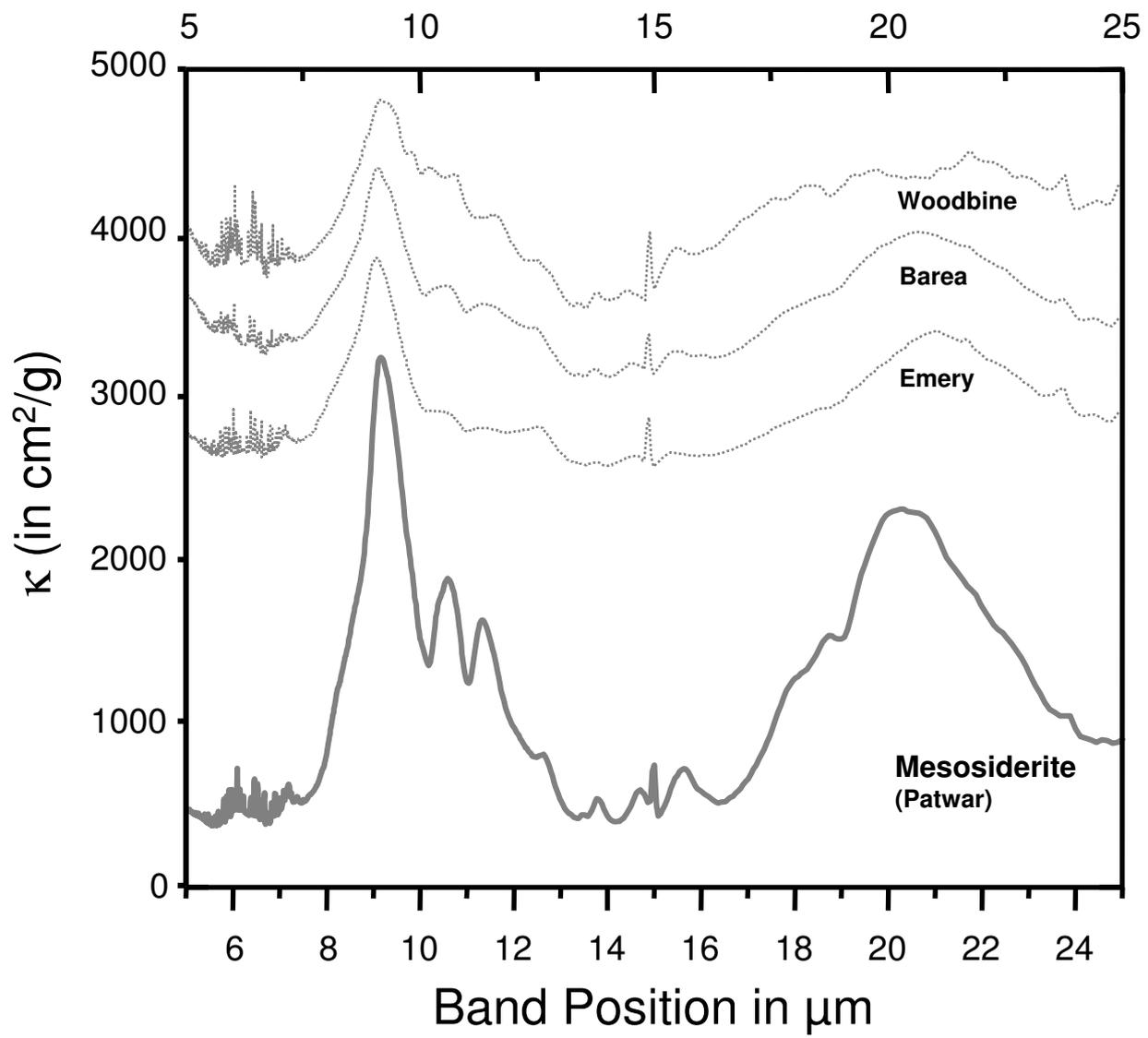

Figure 1c

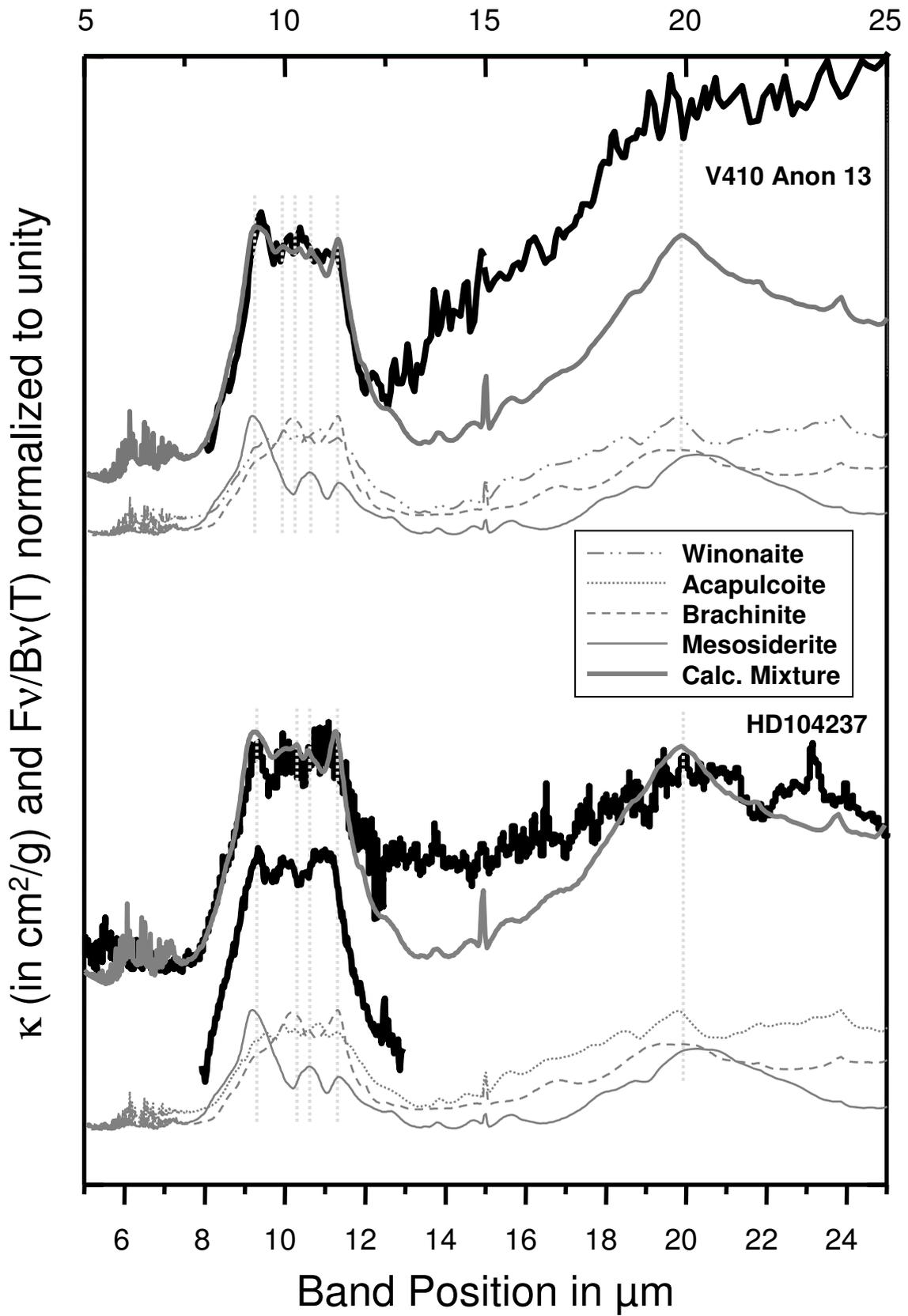

Figure 2a

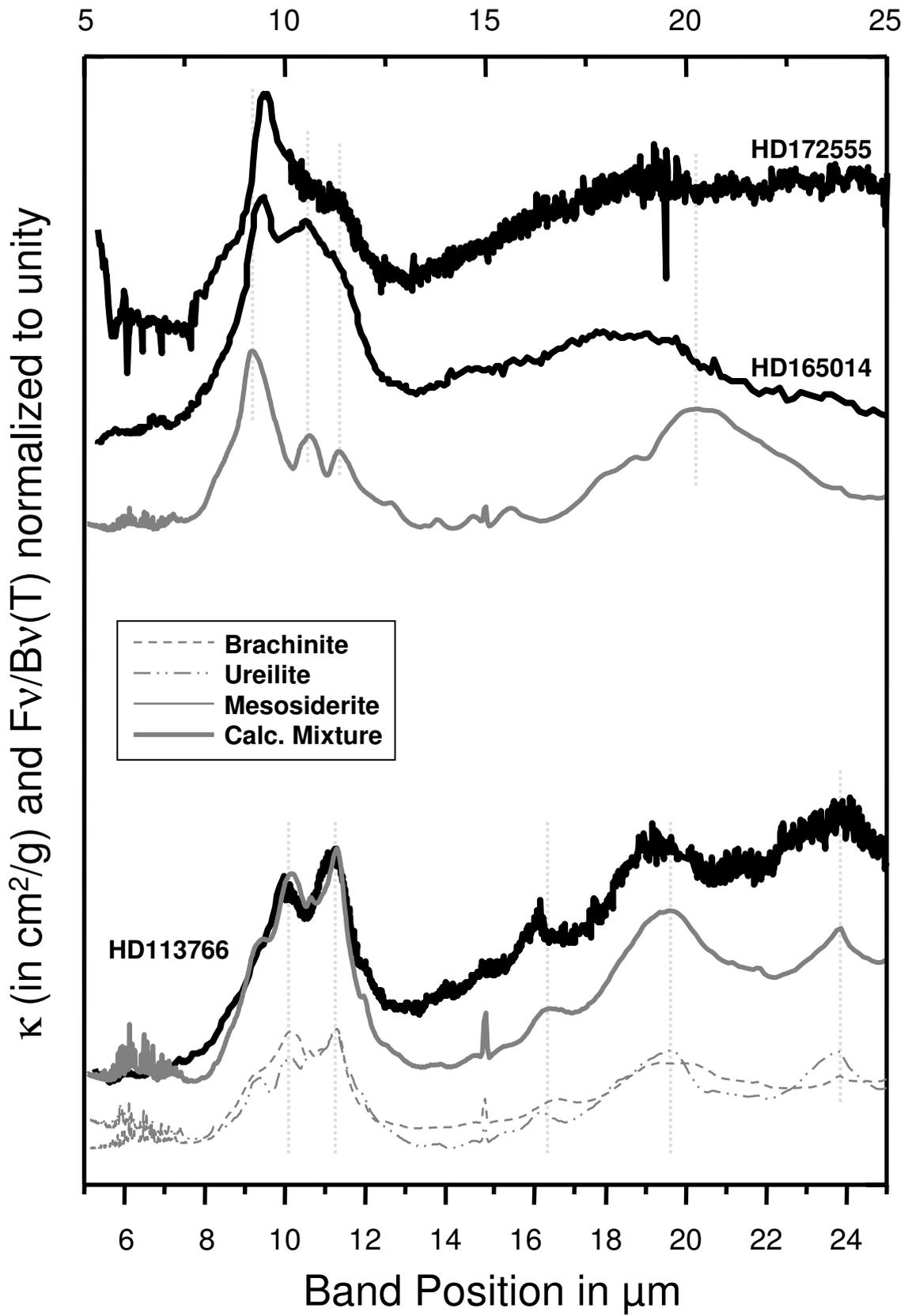

Figure 2b